\documentclass[aps,twocolumn,10pt
nofootinbib,
 amsmath,amssymb,w
 aps,
pre,
]{revtex4-1}
\pdfoutput=1
\usepackage[utf8]{inputenc}
\usepackage{flushend}
\usepackage{dcolumn}
\usepackage{bm}
\usepackage{balance}

\usepackage[normalem]{ulem}

\usepackage{verbatim}
\usepackage{color,ulem}
\usepackage[english]{babel}

\usepackage{subcaption}
\usepackage{graphicx}

\usepackage[colorlinks = true,
            linkcolor = blue,
            urlcolor  = blue,
            citecolor =green,
            anchorcolor = blue]{hyperref}

\usepackage[utf8]{inputenc}
\input Starburst.fd
\newcommand*\initfamily{\usefont{U}{Starburst}{xl}{n}}\initfamily

\newcommand{\beq}{\begin{eqnarray}}
\newcommand{\eeq}{\end{eqnarray}}
\usepackage{amsmath}
\usepackage{tikz}
\usetikzlibrary{decorations.pathmorphing}
\usetikzlibrary{shapes.misc}
\tikzset{cross/.style={cross out, draw=black, minimum size=8*(#1-\pgflinewidth), inner sep=0pt, outer sep=0pt},
cross/.default={1pt}}
\usetikzlibrary{patterns,math}

\begin{document}

\title{Open quantum systems with particle and bath driven by time-dependent fields}

\author{Daniele Gamba$^{1,3}$, Bingyu Cui$^{2}$, Alessio Zaccone$^{1,3}$}
\email{alessio.zaccone@unimi.it}\email{daniele.gamba@theorie.physik.uni-goettingen.de}
\affiliation{${}^1$Department of Physics ``A. Pontremoli'', University of Milan, via Celoria 16,
20133 Milan, Italy}
\affiliation{${}^2$School of Science and Engineering, The Chinese University of Hong Kong, Shenzhen, Guangdong, 518172, P. R. China}
\affiliation{${}^3$Institute for Theoretical Physics, University of G\"ottingen, Friedrich-Hund-Platz 1, 37077 G\"ottingen, Germany.}
\date{\today}

\begin{abstract}
We derive a generalized quantum Langevin equation and its fluctuation-dissipation relation describing the quantum dynamics of a tagged particle interacting with a medium (environment), where both the particle and the environment are driven by an external time-dependent (e.g. oscillating) field. We specialize on the case of a charged tagged particle interacting with a bath of charged oscillators, under an external AC electric field, although the results are much more general and can be applied to any type of external time-dependent fields. We derive the corresponding quantum Langevin equation, which obeys a modified fluctuation-dissipation relation (FDR) where the AC field plays an explicit role. The modified FDR is non-Markovian even if the undriven particle-bath system is Markovian without the external field. We provide an illustration of the usefulness of these results and derive a new form of the quantum Nyquist noise for the voltage fluctuations in electrical circuits under AC conditions (finite frequency), which is the most general since it also accounts for the response of the heat bath (e.g. lattice ions) to the applied AC electric field in the GHz-THz region, of relevance for 5G/6G wireless technologies. This generalized quantum fluctuation-dissipation relation for driven systems can also find other applications ranging from quantum noise in quantum optics to quantum computing with trapped ions.
\end{abstract}

\maketitle

\section{Introduction}
The main goal of the theory of open quantum systems is to understand the situation in which a quantum system of interest interacts with other degrees of freedom \cite{Breuer}. This problem is central to the quantum theory of measurement \cite{Barbatti}, where the manipulation of the quantum system in order to produce an observation implies, necessarily, the interaction of the quantum system with many degrees of freedom (the environment, which includes the measuring apparatus and the observer). Hence, this is a crucial step for the most foundational problems of quantum mechanics, including decoherence \cite{Schlosshauer,Vacchini}, of great relevance for quantum computing, the wavefunction collapse \cite{GRW}, and the quantum-to-classical transition \cite{Ghirardi_book}.
It is also a crucial question for the understanding of how energy dissipation and irreversibility arise in systems that are not at thermodynamic equilibrium. Indeed, historically, a fundamental question of how a quantum system dissipates energy (e.g. the damped quantum harmonic oscillator), suggested by E. Fermi to his student P. Caldirola, marks the beginning of the theory of quantum dissipative systems \cite{Caldirola}. 

One of the most versatile methods to describe a quantum system coupled to an environment, is the Lindblad master equation for the density matrix \cite{Lindblad1976}. The advantage of this approach is that it can work in both the Schr{\"o}dinger and the Heisenberg pictures, and can easily implement continuous symmetry groups. Its main disadvantage is that it is intrinsically Markovian, by construction.

The alternative approach to describe open quantum systems is the Caldeira-Leggett (or Zwanzig) particle-bath model. This leads to the quantum Langevin equation, which is the Heisenberg equation of motion for the (operator) coordinate of a quantum Brownian particle coupled to a heat bath. While the quantum Langevin equation is less versatile than the Lindbladian, e.g. it can only work within the Heisenberg picture, it can naturally accommodate non-Markovianity i.e. history-dependence, which is a hallmark of nonequilibrium dissipative systems. Furthermore, the quantum Langevin equation can be extremely useful in condensed matter and chemical physics to provide single-particle equations of motion which take the complexity of the atomic and molecular environment into account \cite{Ceotto2017,Nitzan}.

The quantum Langevin equation was derived for the first time by Ford, Kac and Mazur \cite{Ford_1965}. Its modern derivation is based on the Caldeira-Leggett (CL) particle-bath model \cite{Ford1987,Ford1988,Weiss2011} and leads to the following (Heisenberg) equation of motion for the coordinate operator $x$
\begin{align}
    m \frac{d^2 x (t)}{d t^2}&= - \frac{\partial V(x)}{\partial {x}}+  {\xi (t)} - \int_{-\infty}^t K(t-t')  \frac{d {x} (t')}{d t'} d t' , \label{QLE}
\end{align}
which is, indeed, the quantum Langevin equation \cite{Weiss2011}.
Here, $K$ is the memory function (or friction kernel), which also gives the (symmetric) correlation for the random force operator $\xi$:
\begin{equation}
\langle [\xi (t),  \xi (t')]_{+} \rangle 
 = \sum_{\alpha} \frac{\hbar m_{\alpha} \nu_{\alpha}^4}{\omega_{\alpha}}
  \coth \left(\frac{\hbar \omega_{\alpha}}{2 k_B T}\right)\cos(\omega_\alpha(t-t')), \label{FDT}
\end{equation}
which has a vanishing mean value:
\begin{equation}
    \langle \xi(t) \rangle =0.
\end{equation}
In the above formulae, the mean anticommutator has been used \cite{Gardiner}:
\begin{equation}
  \langle [\xi (t),  \xi (t')]_{+} \rangle\equiv  \frac{1}{2}\langle  \xi (t)  \xi (t') + \xi (t')  \xi (t)
  \rangle .  
\end{equation}

Furthermore, $V(x)$ is a conservative potential, $m$ is the mass of the quantum particle, $\alpha$ labels the harmonic oscillators of the heat bath, $m_\alpha$ and $\omega_\alpha$ are the mass and frequency of the $\alpha$-th oscillator of the bath, $\nu_{\alpha}$ are the particle-bath coupling coefficients (see below), $k_B$ is the Boltzmann constant and $T$ the temperature.

In the original derivation, there is no external driving force that acts on the particle-bath system as a whole. Previous work has considered the generalization of Eq. \eqref{QLE} to include external force fields. For example, Ref. \cite{Gupta} extended the CL model in a formulation where the coupling is via the momentum variables, to describe a quantum particle coupled to the environment, both of which are subject to an external static magnetic field. While the memory function becomes dependent on the magnetic field, its qualitative form does not change with respect to the field-free case. Other works have considered the effect of an external time-dependent field, such as an electromagnetic field, as acting on the tagged particle only, but not on the bath \cite{Antonenko}. In these cases, a term akin to the classical Lorentz force appears in the quantum Langevin equation (QLE) but the fluctuation-dissipation theorem, Eq. \eqref{FDT}, does not change its form. The issue of a charged particle interacting with both an environment and an external electromagnetic field arises in many areas of physics, from Landau diamagnetism \cite{Landau1930} to the laser cooling and trapping of ultracold atoms and ions \cite{Ming-Gen,Eschner}, of relevance also for quantum computation \cite{Cirac}, to the quantum Hall effect \cite{Klitzing}, 2D electronic materials \cite{Nazarov}, quantum noise in quantum optics \cite{Marquardt}. 

In most of these problems, it is not only the tagged particle to be driven by the external time-dependent field, but also the environment, or thermal bath, in which the particle is embedded. This is clearly not a physically realistic assumption for many systems, where the particle is embedded in the heat bath, and both (the particle and the bath) are subjected to the external time-dependent force field.

The situation, to the best of our knowledge, has been considered only in very few previous contributions, i.e. in Ref. \cite{CuiZaccone2018} and in Refs. \cite{Grabert2018,Indians}. Recently, in Ref. \cite{Albarelli}, a dependence on time has been considered in the coupling part of the particle-bath Hamiltonian. In Ref. \cite{CuiZaccone2018} it was shown, for the classical generalized Langevin equation (GLE), that the time-dependent driving force radically modifies the fluctuation-dissipation relation, introducing a new term to it, proportional to the time autocorrelation of the external field. This result implies that such driven systems cannot, in any instance, be reduced to a Markovian dynamics, and are, thus, intrinsically non-Markovian, due to this new term in the fluctuation-dissipation relation (FDR). It was subsequently shown, in Ref. \cite{Gamba2024}, that this form of the FDR is responsible for strongly anomalous diffusion behaviour in systems of classical charged particles under an external time-dependent electric field, leading to the prediction of hyperballistic superdiffusion regime in the particle transport.

Here, we extend the classical derivation of the generalized Langevin equation for particle-bath systems driven by a time-dependent external field to the quantum regime. Besides the quantum GLE for time-driven open quantum systems, we derive the corresponding FDR, which, also in this case, has an additional term that depends on the time autocorrelation of the external force operator. This term renders the FDR intrinsically non-Markovian, even if the original undriven system is Markovian in the absence of the external time-dependent field.

This result is then used to obtain a new generalized form for the quantum Johnson-Nyquist noise \cite{Callen1951}, that accounts for the effects of the time-dependent driving force on both the "particles" (e.g. electrons) and on the heat bath (e.g. the lattice ions).

\section{The undriven Caldeira-Leggett model}
\subsection{The particle-bath model}
We start with the quantum Caldeira-Leggett model by considering a quantum test particle moving in a generic potential field $V (x)$ and coupled to a bath of quantum harmonic oscillators: 
\begin{equation}
    H_{CL}=H_S+H_B+H_{SB}\label{particlebath}
\end{equation}
with 
\begin{equation}
    H_S=\frac{p^2}{2m}+V(x)
\end{equation}
the Hamiltonian of the particle of mass $m$,
\begin{equation}
    H_B=\frac{1}{2}\sum_\alpha\left(\frac{p_\alpha^2}{m_\alpha}+m_\alpha\omega_\alpha^2 x_\alpha^2\right)
\end{equation}
describing a bath of harmonic oscillators carrying masses $m_\alpha$ and oscillatory frequencies $\omega_\alpha$
and 
\begin{equation}
    H_{SB}=-\sum_\alpha m_\alpha\,\frac{\nu_\alpha^2 }{\omega_\alpha ^2}x\,x_\alpha
\end{equation} the interaction between the system coordinate ($x$) and bath oscillator coordinate ($x_{\alpha}$) with coupling strengths $\nu_\alpha$. For the latter we follow the notation of Ref. \cite{Gamba2024} and define them in units of frequency, unlike other treatments where they are defined in units of frequency squared (e.g. in \cite{Zwanzig2001}), which, however, has obviously no consequences on the final results.  In the above equations, we follow the convention of Ref. \cite{Ford1988}, and we drop hats on all operators and observables of physical quantities. It remains implied that $x$, $p$, $x_{\alpha}$, $p_{\alpha}$ are all Heisenberg operators obeying the usual commutation rules:
\begin{eqnarray}
    [x,p]=i\hbar, \,\,\,\,[x_{\alpha},p_{\beta}]=i\delta_{\alpha\beta}\hbar,
\end{eqnarray}\begin{eqnarray}
   [x,x]=[p,p]=[x_\alpha,x_\beta]=[p_\alpha,p_\beta]=0,
\end{eqnarray}
\begin{eqnarray}
   [x,x_\alpha]=[p,p_\alpha]=[x,p_\alpha]=[p,x_\alpha]=0.
\end{eqnarray}
The bath and interaction terms can be conveniently rearranged as follows:
\begin{eqnarray}
H_B+H_{SB}&=&\frac{1}{2}\sum_\alpha\left(\frac{p_\alpha^2}{m_\alpha}+m_\alpha\omega_\alpha^2\left(x_\alpha-\frac{\nu_\alpha^2}{\omega_\alpha^2}x\right)^2\right) \notag\\ &&-\frac{1}{2}\sum_\alpha m_\alpha\frac{\nu_{\alpha}^4}{\omega_\alpha^2}x^2
  \label{hamiltonianBathInteraction}
\end{eqnarray}
 where the last term is a counter-term representing an initial shift of bath modes \cite{Grabert2018,Vacchini_Petruccione}. 
 
Under an external electric field, the Caldeira-Leggett system is supplemented  by an additional term representing the external time-dependent electric force, 
\begin{equation}
   H_{\text{ext},S}=-q E(t)x,
\end{equation}
where $q$ is the tagged particle's electric charge, and $E(t)$ is the time-dependent electric field. We use the notation $H_{\text{ext},S}$ to make it explicit that, in this case, the external field acts only on the tagged particle, i.e. on the system (S), while it does not act on the bath.
Hence, the total Hamiltonian of the system is given by:
\begin{equation}
    H_{T}=H_{S}+H_B+H_{SB}+H_\text{ext}=H_{CL}+H_{\text{ext},S}
\end{equation}
and a generic quantum observable $A$ satisfies the canonical commutation relation with the Hamiltonian operator:
\begin{equation}
    \frac{d A}{dt}=\frac{i}{\hbar}[H_{T},A] \label{canon}
\end{equation}
which gives the Heisenberg equations of motion for all the variables of interest, i.e. coordinates and momenta of the tagged particle and of the bath oscillators.
In turn, the Heisenberg equations of motion are exactly like the classical equations of motion, with the only difference that the variables are time-dependent Heisenberg operators. The same procedure that one uses to derive the classical GLE can be used here, leading to the quantum GLE \cite{Haenggi-Ingold,Haenggi_QLE,Magalinskii,Zwanzig2001}. One could also work in the second quantization formalism and obtain exactly the same result, cf. pp. 44-58 in Ref. \cite{Gardiner}.

Here, a \emph{caveat} is in order due to the semi-classical nature of the external perturbation $H_{\text{ext},S}=-q E(t)x$. Including this term in $H_{T}$, as done above, is standard and meaningful at the level of semi-classical quantum dynamics \cite{Gupta}, where the electric field $E(t)$ is treated as an explicitly time-dependent, c-number function \cite{Landau_quantum}. Although $E(t)$ is not, strictly speaking, an operator, it can still be included in the Heisenberg equation of motion Eq. \eqref{canon} as long as one recognizes that the Hamiltonian is now explicitly time-dependent \cite{Weiss2011,Landau_quantum}, so that Eq. \eqref{canon} becomes \cite{Sakurai}:
\begin{equation}
    \frac{d A}{dt}=\frac{i}{\hbar}[H_{T}(t),A(t)] + \left( \frac{\partial A}{\partial t}\right)_{\text{explicit}}\label{canon2}
\end{equation}
where the last term is identically zero if the (generic) operator $A$ has no explicit dependence on time, i.e. if it is time-independent in the Schr{\"o}dinger picture. 
The time evolution is thus non-unitary (i.e. not generated by a single time-independent operator), but is still well defined \cite{Landau_quantum,Sakurai}. One can take the commutator $[H_T(t),A(t)]$ in the usual way, even though $H_T$ contains the c-number $E(t)$, which can be treated as a scalar prefactor of operators. Furthermore, the Heisenberg evolution operator becomes time-ordered because the Hamiltonians at different times no longer commute. These considerations apply to all the derivations that are reported in what follows, regardless of whether the external time-dependent electric field acts on the tagged particle only or on both the tagged particle and the heat bath.

\subsection{Generalized Langevin equation}\label{subsect}

From the Hamiltonian Eq. \eqref{particlebath}, for the case of an external field acting only on the tagged particle but not on the bath, one derives the Heisenberg equations of motion for the tagged
particle
\begin{equation}
  m\frac{d^2x(t)}{dt^2} + \frac{\partial V(x(t))}{\partial x} = \sum_{\alpha} m_{\alpha} \nu_{\alpha}^2
  x_{\alpha} (t)-q E(t) \label{unouno},
\end{equation}
and for the bath coordinate operators $x_\alpha$,
\begin{equation}
    \frac{d^2 x_\alpha}{d t^2}=-\omega_\alpha^2 x_\alpha+\nu_\alpha^2x\label{eqnos}.
\end{equation}
In \eqref{eqnos}, $x$ appears as a source term for the $x_{\alpha}$. The solution of \eqref{eqnos} is
\begin{equation} x_{\alpha} (t) = x_{\alpha}^{\hom} (t) + x_{\alpha}^{\text{p}} (t)
   \label{unodue} \end{equation}
where
\begin{eqnarray}
  x_{\alpha}^{\hom} (t) & = & x_{\alpha} (t_0) \cos (\omega_a t) +
  \frac{p_{\alpha} (t_0)}{m_{\alpha}}  \frac{\sin (\omega_{\alpha}
  t)}{\omega_{\alpha}} 
\end{eqnarray}
and
\begin{eqnarray}
  x_{\alpha}^p (t) & = & \int_{-\infty}^t \frac{\sin (\omega_{\alpha} (t -
  t'))}{\omega_{\alpha}} \nu_{\alpha}^2 x (t') d t' .
  \end{eqnarray}

Plugging Eq. \eqref{unodue} into Eq. \eqref{unouno}, we have
\begin{align} m\frac{d^2x(t)}{dt^2} + \frac{\partial V(x(t))}{\partial x}&= \sum_{\alpha} m_{\alpha} \nu_{\alpha}^2
   x_{\alpha}^{\hom} (t)  \notag\\&+\sum_{\alpha} m_{\alpha} \nu_{\alpha}^2
   x_{\alpha}^{\text{p}} (t) -q E(t)\label{eqnlangevin}\end{align}
where $x_{\alpha}^{\text{p}}$ depends on $x (t')$ with $t' < t$, so it can be
interpreted as a memory effect. Integrating by parts, the term containing $x_{\alpha}^{\text{p}}$ becomes $-
\frac{1}{m}\int_{t_0}^t K (t - t') p (t') d t'$, where
\begin{equation} K (t) = \theta (t)\sum_{\alpha}m_\alpha
   \frac{\nu_{\alpha}^4}{\omega_{\alpha}^2} \cos (\omega_{\alpha} t)\label{mem2} \end{equation}
is the time-retarded memory kernel of the frictional force, with $\theta(t)$ the Heaviside step function \cite{Gardiner}.

The first term in the RHS of Eq. \eqref{eqnlangevin}, \begin{equation}\xi(t)=\sum_\alpha m_\alpha\nu_\alpha^2x_\alpha^\text{hom}(t)\label{xinoise}\end{equation} depends on the $\{x_\alpha,p_\alpha\}$ at time $t_0$. Upon averaging under the assumption that the bath is initially at thermal equilibrium, \begin{equation}\langle \xi(t) \rangle = 0\label{xinoise0}\end{equation} and $\langle [\xi (t) \xi (t')]_{+} \rangle $ is given by Eq. \eqref{FDT} \cite{Haenggi-Ingold,Ford1988,Gardiner}.

From Eq. \eqref{eqnlangevin}
one retrieves the standard generalized Langevin equation
\begin{align}
    & m\frac{d^2x(t)}{dt^2}+\int_{-\infty}^tK(t-t')\frac{dx(t')}{dt'}dt'+\frac{\partial V(x(t))}{\partial x}   \\
    & \,\,\,=\xi(t)-q E(t)
    \label{eq:testparticle}.
\end{align}
In the classical limit, the noise operator $\xi$ satisfies
\begin{equation}\langle \xi(t)\xi(t')\rangle=mk_BTK(t-t'),\label{eq:genlange}\end{equation}
as one can demonstrate by imposing that the system is initially at equilibrium, and the initial conditions for momenta and coordinates of the bath oscillators are drawn from the Boltzmann distribution \cite{Zwanzig2001}.

Note that both Eq. \eqref{FDT}, in the quantum limit, and Eq. \eqref{eq:genlange} in the classical limit, have the same form as the corresponding forms of the FDR of a free particle, which is because the external time-dependent force acts only on the tagged particle but not on the bath. Indeed, if the external time-dependent force field acts only on the tagged particle, we then retrieve the classical Langevin equation (in the classical limit) and the quantum Langevin equation (in the quantum limit) with the standard forms of fluctuation-dissipation theorem reported, respectively, in \cite{Zwanzig2001} and in \cite{Ford1988,Weiss2011}.

We now pose the question: what happens if the external time-dependent force field acts not only on the tagged particle but also on the bath? 

In the next section this more general model is studied in detail and a mathematical solution is developed which will lead to a modified fluctuation-dissipation theorem in the quantum limit. While the modified FDT has been studied and reported already for the classical case, it has not been derived yet for the quantum case.

\section{The time-driven particle-bath model}

Let us now consider an external time-dependent driving field, which acts on both the tagged particle and on the heat bath:

\begin{equation}
    H_{\text{ext},SB}=-q E(t)x-\sum_\alpha q_\alpha E(t)x_\alpha,
    \label{eq:Hext}
\end{equation}
where $q_\alpha$ are constants representing the effective charge carried by the $\alpha$-th oscillator. Here, we use the notation $H_{\text{ext},SB}$ to make it explicit that, in this case, the external time-dependent field acts on both the system or tagged particle and on the heat bath.


\begin{figure}[ht]
  \begin{center}
    \includegraphics{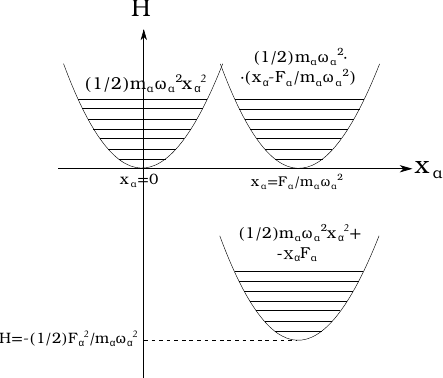} 
  \end{center}
  
  \
  \caption{The $\alpha$-th harmonic well undergoes a shift of $F_{\alpha} /
  (m_{\alpha} \omega_{\alpha}^2)$, due to the coupling with the perturbation
  $F_{\alpha}$. The zero-point energy of the $\alpha$-th oscillator is not
  modified by the presence of the perturbation $F_{\alpha}$. Only the rest
  elongation $x_{\alpha}$ of the $\alpha$-th harmonic oscillator is displaced
  by the presence of the perturbation $F_{\alpha}$, while its energy levels are not
  displaced.}\label{figura2}
\end{figure}


We can describe the driven bath under the effect of the external field via the Hamiltonian Eq. \eqref{hamiltonianBathInteraction}:
\begin{eqnarray}
  H_B+H_{SB}&=&\frac{1}{2}\sum_\alpha\left(\frac{p_\alpha^2}{m_\alpha}+
  m_\alpha\omega_\alpha^2\left(x_\alpha-\frac{F_\alpha}{m_\alpha\omega_\alpha^2}\right)^2\right) \notag\\ &&-\frac{1}{2}\sum_\alpha \frac{F_\alpha^2}{m_\alpha\omega_\alpha^2}\label{modified}
\end{eqnarray}
where now \begin{eqnarray}F_\alpha(t)=m_\alpha\nu_\alpha^2x(t)+q_\alpha E(t) \label{trenda}\end{eqnarray} is the force operator acting on the $\alpha$-th oscillator.
See Fig.\,
\ref{figura2} for the intuitive meaning of the last term ("counter-term") in the RHS of Eq. \eqref{modified}.

Here, the time-dependent electric field $E(t)$ is implemented by the following protocol:
\begin{eqnarray}
  E (t) = \left\{\begin{array}{ll}
    0, & t < 0\\
    E_0 \sin (\Omega t), & t \geq 0
  \end{array}\right.  \label{eq:caseOne}\label{eq:protocol}
\end{eqnarray}
where $E_0$ is the magnitude and $\Omega$ is the loading frequency. 

The total Hamiltonian, in this case, is thus given by:
\begin{equation}
H_{T}=H_{S}+H_B+H_{SB}+H_\text{ext}=H_{CL}+H_{\text{ext},SB} \label{total_driven}
\end{equation}
with the various terms given by the above equations, and where $H_\text{ext}$ is an external field.
From this total Hamiltonian Eq. \eqref{total_driven} and the Heisenberg equations of motion \eqref{canon}, following the same steps as before, \emph{mutatis mutandis}, we derive the following generalized Langevin equation (GLE):

\begin{align}
    m \frac{d^2 x (t)}{d t^2}+\frac{\partial V(x(t))}{\partial x}&=   q E (t) +
  {\eta (t)} \notag\\
   & - \int_{-\infty}^t K(t-t')  \frac{d x (t')}{d t'} d t' 
  \label{eq:langevin2} 
\end{align}
where the new effective noise operator $\eta$ absorbs additional contributions from the charged bath oscillators
\begin{equation}
    \eta (t) = \xi (t) -  \int_0^t M(t -
   t') \frac{d E (t')}{d t'} d t'.
   \label{eq:etat}
\end{equation}
The second term in the right hand side of Eq. \eqref{eq:etat} reflects the effect of the field on the test particle via its coupling to the surrounding bath oscillators, which themselves respond to the AC electric field. As we will elucidate explicitly in the next section, such effect amounts to a new, additional term in the Nyquist theorem for quantum noise. We also note that, as usual, the memory function in the damping term, the last term on the right hand side of Eq. \eqref{eq:langevin2}, has the same form given in Eq. \eqref{mem2}.

In the above Eq. \eqref{eq:etat}, $M$ is the force delay kernel,
\begin{equation} M (t) = \sum_{\alpha} q_{\alpha} \frac{\nu_{\alpha}^2}{\omega_{\alpha}^2}
   \cos (\omega_{\alpha} t) ,\label{mem}\end{equation}
   which is independent of temperature and is only a function of the bath oscillators parameters. This kernel incorporates the spectral density of the thermal bath oscillators and their coupling to the external field. It describes the additional effect of the bath on the noise  accounting for the fact that the bath statistics is altered due to its being driven by the external field. These results are consistent, apart from differences in the notation, with previous derivations \cite{Grabert2018,CuiZaccone2018}.
If we take the continuous limit for the sum over oscillators $\alpha$, we can write Eq. \eqref{mem2} and Eq. \eqref{mem} as integrals over a (e.g. Debye) density of states of the set of bosonic oscillators \cite{Zwanzig2001,Gardiner,CuiZaccone2018}. 


\section{Fluctuation-dissipation theorem with the driven bath}
For ergodic systems, there is no difference between the ensemble average and the long time average of observables. However, when the system is subject to an external driving, as in our case, the ergodicity usually breaks down. We thus look at the time average of an operator $A$,
\begin{equation}
  \langle A (t) \rangle = \frac{1}{T}  \int_t^{t + T} A (t') dt' 
  \label{eq:timeAverage}
\end{equation}
and its self-correlation 
\begin{equation}
  \langle A (t) A (t + \tau) \rangle = \frac{1}{T}  \int_t^{t
  + T} A (t') A  (t' + \tau) dt' .
  \label{eq:timeAverage2}
\end{equation}

The bath is at thermal equilibrium before the external driving field is switched on at $t=0$. Assuming that the equilibrium system (for $t \leq 0)$ is ergodic, the time average in Eq. \eqref{eq:timeAverage} is equal to the following ensemble average (cf. page 103 in Ref. \cite{Zwanzig2001}): 
\begin{equation}
     \langle A(t) \rangle = \frac{\text{Tr} \left[A (t) e^{-
   H_{B} / k_B T}\right]}{Z}\label{average22}
\end{equation}
with
\begin{equation}
    Z = \text{Tr} \left[e^{- H_{B} / k_B T}\right],
\end{equation}
where we used the equilibrium density matrix defined as:
\begin{equation}
    \rho_{eq}=\frac{1}{Z}e^{-H_B/k_B T},
\end{equation}
which is itself a quantum operator in the Heisenberg picture \cite{Zwanzig2001}.

Upon averaging Eq. \eqref{xinoise} (the noise operator), we get zero as in Eq. \eqref{xinoise0}: the average noise of an equilibrium system vanishes, as expected.
The symmetric self-correlation of the random force operator acting on the quantum tagged particle, for the case where the external time-dependent field does not act on the bath, is 
\begin{align}
&C(t,t')\equiv \langle [\xi (t) , \xi (t')]_{+} \rangle
   =\notag \\
   &\sum_{\alpha} \frac{\hbar m_{\alpha} \nu_{\alpha}^4}{\omega_{\alpha}}
  (2  \bar{n}_{\alpha} + 1)\text{cos}(\omega_\alpha(t-t')), \label{eq;noiseCorrelation} 
\end{align}
where $\bar{n}_{\alpha}$ is the Bose-Einstein occupation factor, evaluated at $t=0$ (when the bath is still at equilibrium). We then have
\begin{equation}
 C(t,t')
   =\sum_{\alpha} \frac{\hbar m_{\alpha} \nu_{\alpha}^4}{\omega_{\alpha}}
  \coth \left(\frac{\hbar \omega_{\alpha}}{2 k_B T}\right)\cos(\omega_\alpha(t-t')).
  \label{eq:FDT}
\end{equation} Eq. \eqref{eq:FDT} is shown in Appendix \ref{appendice}.

It is interesting to note that, at high temperature, $\coth(x)\sim\mathcal{O}(1/x)$, so, in that limit ($k_B T \gg \hbar \omega$) one recovers the classical result \cite{Zwanzig2001}:
\begin{equation}
    C(t,t')
   = \sum_\alpha\frac{2k_BTm_\alpha\nu_{\alpha}^4}{\omega_\alpha^2}\text{cos}(\omega_\alpha(t-t')).
\end{equation}

Equation \eqref{eq:FDT} describes the autocorrelation of the noise, $\xi$. However, in the presence of a dynamical coupling between the external driving and the bath, the effective noise on the tagged particle is given by Eq. \eqref{eq:etat}. Since the bath oscillators respond to the driving field, cf. Eq. \eqref{eq:Hext}, such effect must be accounted for explicitly, and will lead to a modification of the canonical FDR, similar to what happens in the classical limit \cite{CuiZaccone2018}. 
Indeed, when an AC electric field is applied to both the particle and the bath, the average of Eq. \eqref{eq:etat} incorporates the additional field effect on the noise. This average is computed under the assumption that the system is at equilibrium at $t\leq 0$. Shortly afterward at $t \approx 0$, while the system remains near equilibrium, we evaluate the average of Eq. \eqref{eq:etat} using Eq. \eqref{average22} with $H_B=H_B(t=0)$, following the approach in Refs. \cite{Zwanzig2001} and \cite{Gamba2024}. This procedure yields the correlation function
\begin{eqnarray}
&&   \langle [\eta (t) , \eta (t')]_{+} \rangle =  C(t,t') +\\& +&
  \left( \int_0^t M (t - s) \frac{d E (s)}{d s} d s \right) \notag\\
  & \times  &\left( \int_0^{t'} M (t' - s) \frac{d E (s)}{d s} d s \right)\notag\\
  & = & \sum_{\alpha} \frac{\hbar m_{\alpha} \nu_{\alpha}^4}{\omega_{\alpha}}
 \coth \left(\frac{\hbar \omega_{\alpha}}{2 k_B T}\right)\text{cos}(\omega_\alpha(t-t')) +\notag\\
  & +  & \left( \int_0^t M (t - s) \frac{d E (s)}{d s} d s
  \right) \notag\\
  & \times  &  \left( \int_0^{t'} M (t' - s) \frac{d E (s)}{d s} d s
  \right) \label{newFDT}
\end{eqnarray}
where $t$ and $t'$ are defined on the positive real semi-axis.
Equation \eqref{newFDT} is a new form of the quantum fluctuation-dissipation theorem (FDT) valid whenever both the tagged particle and the bath are driven by a time-dependent field. 
Comparing Eq. \eqref{newFDT} to the equilibrium form of the quantum FDT \cite{Ford1988}, we notice the presence of a new correlation term, the second term on the right hand side of Eq. \eqref{newFDT}. 

This new nonequilibrium correction term depends on the external electric field $E(t)$, which is a c-number function of time, not a quantum operator or observable. We should recall the \emph{caveat} at the end of Section II.A: $E(t)$ does not belong to the system's Hilbert space. It does not evolve under Heisenberg time evolution, and it does not satisfy canonical commutation relations. But, the total Hamiltonian $H_T$, which includes $E(t)$, still satisfies commutation relations with any generic quantum observable $A$ according to Eq. \eqref{canon2}. Furthermore, the evolution is time-ordered. The role of $E(t)$ is akin to a background input or an external control parameter, and this is a valid and common treatment in linear response theory \cite{kapusta_gale_2023}, Keldysh theory \cite{Kamenev_2011} and open quantum systems \cite{Breuer}.

Hence, the second term appearing in the right hand side of Eq. \eqref{newFDT} reflects higher-order response or modifications to the fluctuation spectrum due to external driving of the thermal bath, and we shall look at it more closely, along with its physical implications, in the next section.


\section{Generalized quantum Nyquist noise}

Following {\cite{Gardiner,Gaspard2022}}, we can use the above results to infer a fluctuation-dissipation relation for the random fluctuations of the voltage in a circuit, in the quantum regime. In
Fourier space, for an unconstrained particle, Eq. \eqref{eq:langevin2} becomes \cite{Gardiner}
\begin{equation}
  [-m \omega^2 + i\omega K (\omega)] x (\omega) = qE (\omega) + \eta (\omega),
  \label{eq:chargemotion}
\end{equation}
where $\omega$ is a generic driving frequency. 
Following Ref. \cite{Gaspard2022}, we divide Eq. \eqref{eq:chargemotion} through by a uniform charge density parameter, $\lambda = qnA$, where $q$ is the elementary charge, $n$ the medium's density of charge carriers, $A$ the cross-section of the circuit. Then we introduce the following quantities: the electric current $I=\lambda \dot{x}$, induction $L=m/\lambda^2$ and alternating voltage $V (\omega) = qE (\omega)/\lambda$ with random fluctuations $V_I (\omega) = \eta/\lambda$. We also define the frequency-dependent resistance \begin{equation}R (\omega) =
m K (\omega)/\lambda^2.\end{equation}
This turns Eq. \eqref{eq:chargemotion} into
\begin{equation}
  i \omega LI (\omega) + R (\omega) I (\omega) = V (\omega) + V_I (\omega).
  \label{eq:circuitmacro}
\end{equation}
Equation \eqref{eq:circuitmacro} describes the current in an LR circuit subjected to an AC voltage, which crucially includes new contributions to the fluctuations because of the response of the charged heat bath environment to the external voltage.

The role of the random voltage fluctuations $V_I$ is analogous to that of the random force $\eta$ in Eq. \eqref{eq:etat}, and it, too, accounts for the additional effect of the external driving field on the bath statistics. As a matter of fact, the charge carriers flowing through a resistor are subject to random noise due to the continuous thermal motion of the surrounding charges (the ions of the lattice), which constitute the "charged oscillators" of the thermal bath, in our model.

Upon introducing the above described quantities, Eq. \eqref{newFDT} takes the form
\begin{align}
   & \langle V_I (t) V_I (t') \rangle = \sum_{\alpha} m_{\alpha} \hbar \omega_{\alpha} R  (\omega_{\alpha}) f
(\omega_{\alpha})  \text{cos} (\omega_{\alpha} (t - t'))  \notag\\ + 
  &\frac{1}{q^2} \left( \int_0^t M (t - s) \frac{dV (s)}{ds} ds
  \right)\left( \int_0^{t'} M (t' - s) \frac{dV (s)}{ds} ds \right)
   \label{eq:sedici}
\end{align}
which generalizes the conventional form of the Johnson-Nyquist theorem {\cite{Johnson1928,Nyquist1928, Kittel1959, Lawson1950}} and is valid for $t>0$. Here, $f(\omega_{\alpha})\equiv \coth (\hbar \omega_{\alpha} / 2 k_B T)$. In particular, the second term on the right hand side of Eq. \eqref{eq:sedici} is new (it is not present in the standard Johnson-Nyquist form of the FDR) and is caused by the response of the charged bath to the applied voltage. In the classical limit,  setting $V=0$, and defining $R = \frac{1}{2} \sum_{\alpha} m_{\alpha} R 
(\omega_{\alpha})$, by following similar steps as in \cite{Gardiner}, pp. 48-49, we recover the standard Johnson-Nyquist theorem.

To get a further insights into the effect of field-responsive bath statistics on the voltage fluctuation $\langle V_I(t)V_I(t+ \tau)\rangle$, following
Refs. {\cite{Zhitlukhina2023,Buttiker1998}}, we define the noise spectrum as the
average
\begin{equation}
  S (\tau) = \frac{\Delta f}{4 \pi}  \int_0^{2 \pi / \Delta f} \langle V_I (t +
  \tau) V_I (t) \rangle dt
\end{equation}
where $\Delta f$ is the frequency bandwidth of the circuit. In this definition, the spectral density $S(\tau)$ has units of [volts]$^2$. Taking the integral over a finite time window equal to the inverse of the frequency resolution is a common trick in signal processing to estimate correlations without needing to go to infinite time \cite{Gardner}. 

Combining this with Eq.
\eqref{eq:sedici}, the noise spectrum is

\begin{align}
   & S(\tau) = \sum_{\alpha} m_{\alpha} \hbar \omega_{\alpha} R  (\omega_{\alpha}) f
(\omega_{\alpha})  \text{cos} (\omega_{\alpha} \tau)  +\notag\\ + 
  &\frac{1}{q^2}\frac{\Delta f}{4 \pi}  \int_0^{2 \pi / \Delta f}\text{d}t \left( \int_0^{t+\tau} M (t+\tau - s) \frac{dV (s)}{ds} ds
  \right)\notag\\&\times\left( \int_0^{t} M (t - s) \frac{dV (s)}{ds} ds \right).
   \label{eq:diciotto}
\end{align}
Now we apply the following initial conditions for Eq. \eqref{eq:chargemotion} in the form of a protocol for $V(t)$ being switched on at $t=0$, similar to Eq. \eqref{eq:protocol},
\begin{equation}
  V (t) = \frac{qE_0}{\lambda} 
    \sin (\Omega t) \theta (t).
 \label{eq:field19}
\end{equation}
Upon applying this protocol for the externally applied voltage, we obtain the following form of the fluctuations spectrum:
\begin{align}
   & S(\tau) = \sum_{\alpha} m_{\alpha} \hbar \omega_{\alpha} R  (\omega_{\alpha}) f
(\omega_{\alpha})  \text{cos} (\omega_{\alpha} \tau) + \notag\\ + 
  &\frac{\Delta f E_0^2}{4 \pi\lambda^2}  \int_0^{2 \pi / \Delta f}\text{d}t \left( \int_0^{t+\tau} \Omega\, M (t+\tau - s) \text{cos}(\Omega s) ds
  \right)\notag\\&\times\left( \int_0^{t} \Omega\, M (t - s) \text{cos}(\Omega s) ds \right).
  \label{eq:Stau}
\end{align}

This equation represents a generalized Johnson-Nyquist relation valid in the quantum regime (and of course also in the classical regime upon taking the relevant limit). It describes a system where not only the electrons, but also the ions of the lattice of the conductor, which form the heat bath, respond to the applied AC voltage. At frequencies used in standard electrical and electronic circuits, the actual values of electric field inside the conductors are very low and the motion of the ions in response to the applied AC field is negligible. However, at very high frequencies, GHz and especially THz, the dynamics of the lattice ions will be affected by the AC field. In that regime, we expect that this new term in the fluctuation-dissipation relation may be an important contribution to the noise.

The first term on the right hand side of Eq. \eqref{eq:Stau} coincides with the standard fluctuation-dissipation relation of quantum systems \cite{Weiss2011}: by following standard steps that can be found in the textbooks (cf. pages 60-61 in \cite{Gardiner}), and assuming a frequency-independent damping (resistance) $R$, it recovers the classical limit of the standard Johnson-Nyquist noise, $S(\tau)/\Delta f \rightarrow 4\, R\, k_B T$, where $\Delta f$ is the bandwidth. The second term on the right hand side of Eq. \eqref{eq:Stau} is new, and has been derived here for the first time.
This is an additional contribution to the expected spectral amplitude, which can be observed experimentally in AC circuits. Importantly, a fundamental implication of this result is that, due to the field-dependent new term on the right hand side of Eq. \eqref{eq:Stau}, the noise spectrum will typically be non-Markovian, i.e. it will depend on the frequency of the driving field. This feature has been observed for example in a recent realistic model of quantum transport in AC fields in Ref. 
\cite{Zhitlukhina2023}. 
In the next section we shall analyze this correction term in more detail.

\section{Driven-bath contribution to the Nyquist noise at high frequencies}
We shall now analyze the contribution of the AC-driven bath (crystal lattice ions) to the spectral density of the noise. We focus on the second term on the right hand side of Eq. \eqref{eq:Stau}, and we first solve the integrals analytically, by recalling the form of the force delay kernel, $M(t)$, from Eq. \eqref{mem}. We obtain:
\begin{align}
   &\left( \int_0^{t}  M (t - s) \text{cos}(\Omega s) ds \right)\\
  &=\sum_{\alpha} q_{\alpha} \frac{\nu_{\alpha}^2}{\omega_{\alpha}^2} \int_{0}^{t}\cos[\omega_\alpha(t-s)\cos(\Omega s)]ds \notag\\
  & =\sum_{\alpha} q_{\alpha} \frac{\nu_{\alpha}^2}{\omega_{\alpha}^2}\frac{\omega_\alpha  \sin ( \omega_\alpha t)-\Omega  \sin (\Omega t)}{\omega_\alpha ^2-\Omega ^2}
\end{align}
Now we convert the sum over discrete phonon frequencies $\omega_{\alpha}$ to a continuous integral, as is standard (cf. page 23 in \cite{Zwanzig2001}):
\begin{equation}
    \sum_{\alpha} q_{\alpha} \frac{\nu_{\alpha}^2}{\omega_{\alpha}^2} ... \rightarrow \bar{q}\int_{0}^{\omega_D}\rho(\omega)\frac{\nu^2}{\omega^2} ...d\omega \label{Debye}
\end{equation}
where $\omega$ is the phonon frequency and $\rho(\omega)$ is the phonon density of states of the conductor. Furthermore, for simplicity we also took $\nu$, the electron damping rate, to be a constant, again following \cite{Zwanzig2001}, and we took the average charge of the lattice positive ions, $\bar{q}$, out of the integral. This will of course depend on the valence state of the lattice ions, e.g. $\bar{q}$ will be equal to the elementary charge if the ions have valence +1. Also, there is an ultraviolet cutoff in the integral, which is, as usual, given by the Debye frequency of the solid, $\omega_D$, typically in the THz. 
Hence we evaluate 
\begin{equation}
\bar{q}\int_{0}^{\omega_D}\rho(\omega)\frac{\nu^2}{\omega^2} \frac{\omega  \sin ( \omega t)-\Omega  \sin (\Omega t)}{\omega ^2-\Omega ^2}d\omega  \label{integral}
\end{equation}
 For the phonon density of states we take a Debye form: 
\begin{equation}
    \rho(\omega) = A_D\,\omega^2
\end{equation}
where $A_D = 9 /\omega_D^3$ and $\omega_D$ is the Debye frequency of the solid. The above integral \eqref{integral} admits an analytical solution, which is lengthy and contains SineIntegral and CosIntegral functions. The  SineIntegral and the CosIntegral functions, in the solution, contain expressions like $\Omega \pm \omega_D$ as their argument. Considering that $\omega_D$ is a very big number (typically THz) and that the SineIntegral function for large values of its argument can be approximated with $\pi/2$, and also considering that the CosIntegral function, instead, goes to zero for large values of its argument, and that terms like $\log \left(\omega _D+\Omega \right)-\log \left(\omega _D-\Omega \right)$ cancel each other because $\omega_D$ is big compared to $\Omega$, then the result of the integral over $\omega$, barring prefactors, becomes:
\begin{equation}
    \frac{1}{2} \pi  \cos ( \Omega t).
\end{equation}
For the factor $ \left( \int_0^{t+\tau} M (t+\tau - s) \cos(\Omega s) ds
  \right)$ the calculation is exactly the same, with the only difference that we have $t+\tau$ instead of $t$, leading to:
  \begin{equation}
    \frac{1}{2} \pi  \cos ( \Omega (t+\tau)).
\end{equation}
Hence by putting all things together we get, for the driven-bath contribution  $S_{DB}(\tau)$ to the noise spectrum:
\begin{equation}
    S_{DB}(\tau)=\frac{\pi^3}{8} A_D^2 \,\nu^4 \Delta f\, \Omega \cos(\Omega \tau)\,\bar{q}^2\frac{ E_0^2}{\lambda^2},
\end{equation}
where the group $\bar{q}^2\frac{ E_0^2}{\lambda^2}$ has dimension of volts.

We notice the explicit frequency dependence of the noise, which implies that there is intrinsic non-Markovian character to it, as one could expect. Also, this form indicates that the new noise contribution from lattice ions response to the AC field has cyclostationary properties.

We can attempt to estimate the upper bound (when the cosine in the above formula is equal to one) for this new contribution for a copper wire at room temperature. The Debye prefactor for copper is $6.72 \times 10^{-41}$ $s^3$, the characteristic damping rate for electrons in copper is $4 \times 10^{13}$ $s^{-1}$, and we take a THz frequency, $\Omega = 10^{12}$ Hz, and a bandwidth $10^{12}$ Hz. Furthermore, the group $\bar{q}^2\frac{ E_0^2}{\lambda^2}$, which has dimension of volts, can be set equal to the applied voltage in the circuit, which, for THz antennas and waveguides is of the order of $1$ V. Over a time period $T$ that we take much longer than the inverse oscillation frequency, we thus have an average noise:
\begin{align}
    \bar{S}_{DB}(\Omega)&=\frac{\pi^3}{8} A_D^2 \,\nu^4 \Delta f\, \Omega \left(\frac{1}{T}\int_0^{T}\cos(\Omega \tau) d\tau\right)\,\bar{q}^2\frac{ E_0^2}{\lambda^2}\\
    & =\frac{\pi^3}{8} A_D^2 \,\nu^4 \Delta f\, \frac{\sin (\Omega T)}{ T}\,\bar{q}^2\frac{ E_0^2}{\lambda^2}  .
\end{align}
with the above typical values and taking $\frac{\sin (\Omega T)}{ T}$ to be of order one, we thus get a maximum noise
\begin{equation}
    S_{DB} \approx 10^{-11} \, \text{V}^2
\end{equation}
which leads to a root mean noise of $3$ $\mu$V. This is a meaningful correction if compared to typical values of Johnson-Nyquist noise in THz devices, which are in the order of 1 mV.

\section{Conclusions}
In summary, we have studied the Caldeira-Leggett model of quantum dissipation in open quantum systems subjected to an external time-dependent force that acts simultaneously on both the particle (the system) and the heat bath (the environment). This setup is important for both technological applications (from noise in AC-current quantum electronics especially at high frequencies, GHz-THz, to qubits decoherence) and for foundational quantum physics (e.g. the quantum wavefunction collapse problem, where typically the "measurement" is conducted with an optical field). We started from the standard Caldeira-Leggett coupling Hamiltonian and added an external time-dependent (electric) field semi-classically. We thus considered the dissipative dynamics of an electrically-charged tagged particle that interacts with a bath of charged bosonic oscillators, where both the particle and the bosonic oscillators are driven by an external time-dependent (AC) electric field.

A new fluctuation-dissipation relation for the tagged particles interacting with the time-driven bath is derived, Eq. \eqref{newFDT}, which is valid in the quantum regime (low temperature and/or very high frequencies) and reduces to the Johnson-Nyquist noise upon taking the classical limit for a non-driven bath.
We then used this generalized fluctuation-dissipation relation for driven systems to obtain a closed-form expression
for the spectral density of the noise. We attempted to estimate the contribution from the driven-bath to the overall Nyquist noise and found that, for THz devices, this could amount to a few $\mu$V, to be compared with an expected Nyquist noise (which neglects the contribution from the driven bath) of about 1 mV. 
This effect may be important in future quantum technologies operating at high frequencies, as well as for tackling the decoherence problem in presence of optical fields \cite{Eberly}. Moreover, it provides a generalized framework for nonequilibrium quantum systems under external driving, with many potential applications ranging from quantum computing \cite{Olivares} to quantum optics applications \cite{Aspect}.

\section*{Acknowledgements}
A.Z. gratefully acknowledges funding from the European Union through Horizon Europe ERC Grant number: 101043968 ``Multimech'', from US Army Research Office through contract nr. W911NF-22-2-0256, and from the Nieders{\"a}chsische Akademie der Wissenschaften zu G{\"o}ttingen in the frame of the Gauss Professorship program. B.C. acknowledges funding from the National Natural Science Foundation of China (No. 12404232) and the start-up funding from the
Chinese University of Hong Kong, Shenzhen (No. UDF01003468).

\appendix 
\section{Fluctuation-dissipation relation}\label{appendice}
In this Appendix, we briefly recall the derivation of the standard quantum fluctuation-dissipation relation for the Caldeira-Leggett model leading to the quantum Langevin equation, in the limit of an un-driven bath, cf. e.g. Ref. \cite{Gardiner}. This will clarify the meaning of quantities and relations used throughout the main article.

Introduce ladder operators $a_{\alpha}$, $a_{\alpha}^{\dagger}$ for the bath, such that
\begin{equation}
  x_{\alpha} - \frac{F_{\alpha}}{m_{\alpha} \omega_{\alpha}^2} = i \left(
  \frac{\hbar}{2 m_{\alpha} \omega_{\alpha}} \right)^{1 / 2}  (a_{\alpha} (t)
  - a_{\alpha}^{\dagger} (t)),
\end{equation}
\begin{equation}
  p_{\alpha} = \left( \frac{m_{\alpha} \hbar \omega_{\alpha}}{2} \right)^{1 /
  2}  (a_{\alpha} (t) + a_{\alpha}^{\dagger} (t)) .
\end{equation}
with $a_{\alpha} (t) = a_{\alpha} e^{- i \omega_{\alpha} t}$,
$a_{\alpha}^{\dagger} (t) = a_{\alpha}^{\dagger} e^{i \omega_{\alpha} t}$, and
$F_{\alpha}$ given by Eq. \eqref{trenda}. The term $\xi (t)$ appearing in Eq.
\eqref{eq:etat} reads explicitly
\begin{eqnarray}
  \xi (t) & = & \sum_{\alpha} m_{\alpha} \nu_{\alpha}^2   \nonumber\\
  &  &\times\left[ \left(
  x_{\alpha} (0) - \frac{F_{\alpha} (0)}{m_{\alpha} \omega_{\alpha}^2} \right)
  \cos \omega_{\alpha} t + \frac{p_{\alpha} (0)}{m_{\alpha} }  \frac{\sin
  \omega_{\alpha} t}{\omega_{\alpha}} \right] \nonumber\\
  & = & \sum_{\alpha} im_{\alpha} \nu_{\alpha}^2 \left( \frac{\hbar}{2
  m_{\alpha} \omega_{\alpha}} \right)^{1 / 2} \nonumber\\
  &  & \times (\{a_{\alpha} - a_{\alpha}^{\dagger} \} \cos \omega_{\alpha} t
  - i\{a_{\alpha} + a_{\alpha}^{\dagger} \} \sin \omega_{\alpha} t)
  \nonumber\\
  & = & \sum_{\alpha} im_{\alpha} \nu_{\alpha}^2 \left( \frac{\hbar}{2
  m_{\alpha} \omega_{\alpha}} \right)^{1 / 2}  (a_{\alpha} e^{- i
  \omega_{\alpha} t} - a_{\alpha}^{\dagger} e^{i \omega_{\alpha} t}).  \nonumber\\
  &  &
\end{eqnarray}
This quantity $\xi (t)$ depends on the bath phase space $\{a_{\alpha},
a_{\alpha}^{\dagger} \}_{\alpha = 1... N}$ at $t = 0$, and is oscillating with
the bath's characteristic frequencies $\omega_{\alpha}$. We can estimate its
correlations as
\begin{eqnarray}
  \langle \xi (t) \xi (t') \rangle &=& \sum_{\alpha} \frac{\hbar m_{\alpha}
  \nu_{\alpha}^4}{\omega_{\alpha}} e^{- i \omega_{\alpha}  (t - t')} \langle
  \{a_{\alpha}, a_{\alpha}^{\dagger} \} \rangle\nonumber\\
  &=& \sum_{\alpha} \frac{\hbar m_{\alpha} \nu_{\alpha}^4}{\omega_{\alpha}} e^{-
  i \omega_{\alpha}  (t - t')} \langle 2 a_{\alpha}^{\dagger} a_{\alpha} + 1
  \rangle\nonumber\\
  &=& \sum_{\alpha} \frac{\hbar m_{\alpha} \nu_{\alpha}^4}{\omega_{\alpha}} e^{-
  i \omega_{\alpha}  (t - t')}  (2 \langle n_{\alpha} \rangle + 1),\nonumber\\
  \label{eq:5.17}
\end{eqnarray}
with $n_{\alpha} = a_{\alpha}^{\dagger} a_{\alpha}$. Using \eqref{average22},
we have
\begin{eqnarray}
  \langle n_{\alpha} \rangle & \equiv & \langle a_{\alpha}^{\dagger}  (0)
  a_{\alpha} (0) \rangle \nonumber\\
  & = & \frac{1}{Z} \text{Tr} (a_{\alpha}^{\dagger} a_{\alpha} e^{-
  H_{B} / k_B T}) \nonumber\\
  & = & \frac{\sum_{\alpha = 0}^{\infty} n_{\alpha} e^{- \hbar \omega_{\alpha} \left( n +
  \frac{1}{2} \right) / k_B T}}{\sum_{\alpha = 0}^{\infty} e^{- \hbar
  \omega_{\alpha} \left( n + \frac{1}{2} \right) / k_B T}} \nonumber\\
  & = & \frac{\coth (\hbar \omega_{\alpha} / 2 k_B T) - 1}{2},
  \label{treis}
\end{eqnarray}
where $H_B$ is the Hamiltonian of the bath, as introduced in the main article.

Using \eqref{treis} in \eqref{eq:5.17}, and taking the real part,
\begin{eqnarray}
  \langle {\xi} (t) {\xi} (t') \rangle & = & \sum_{\alpha} \frac{\hbar
  m_{\alpha} \nu_{\alpha}^4}{\omega_{\alpha}}  (2 \langle n_{\alpha} \rangle +
  1) \cos (\omega_{\alpha} (t - t'))\nonumber\\
  & = & \sum_{\alpha} \frac{\hbar m_{\alpha} \nu_{\alpha}^4}{\omega_{\alpha}}
  \coth (\hbar \omega_{\alpha} / 2 k_B T) \cos (\omega_{\alpha} (t -
  t')),\nonumber\\
\end{eqnarray}
which is Eq. \eqref{eq:FDT} in the text.

\bibliographystyle{apsrev4-1}
\bibliography{refs}

\end{document}